%% file: wccm2020.tex
\documentclass[a4paper]{wccm2020}
\usepackage[indent=5mm]{parskip}
\usepackage{pslatex}
\usepackage{graphicx}
\usepackage{amsmath}
\usepackage{amsfonts}
\usepackage{mathptmx}
\usepackage{amssymb}
\input{definitions}
\title{Physics-aware, deep probabilistic modeling of multiscale dynamics in the Small Data regime}

\author{Sebastian Kaltenbach$^1$, Phaedon-Stelios Koutsourelakis$^1$}

\heading{Sebastian Kaltenbach, Phaedon-Stelios Koutsourelakis}

\address{$^{1}$ Professorship of Continuum Mechanics, Technical University of Munich\\
Boltzmannstr.15, 85748 Garching\\
\{sebastian.kaltenbach; p.s.koutsourelakis\}@tum.de, www.mw.tum.de/contmech}


\abstract{\textit{The data-based  discovery of effective, coarse-grained (CG) models of high-dimen- sional dynamical systems presents a unique challenge in computational physics and particularly in the context of multiscale problems. The present paper offers a probabilistic perspective that simultaneously identifies predictive, lower-dimensional coarse-grained (CG) variables as well as their dynamics. We make use of the expressive ability of deep neural networks in order to represent the right-hand side of the CG evolution law. Furthermore, we demonstrate how domain knowledge that is very often available in the form of physical constraints (e.g. conservation laws) can be incorporated with the novel concept of virtual observables. Such constraints, apart from leading to physically realistic predictions, can significantly reduce the requisite amount of training data which enables reducing the amount of required, computationally expensive multiscale simulations (Small Data regime). The proposed state-space model is trained using  probabilistic inference tools and, in contrast to several other  techniques,  does not require the prescription of a fine-to-coarse (restriction) projection nor  time-derivatives of the state variables. The formulation adopted is capable of quantifying the predictive uncertainty as well as of reconstructing  the evolution of the full, fine-scale system which allows to select the quantities of interest a posteriori. We  demonstrate the efficacy of the proposed framework in a high-dimensional system of moving particles.}}

\begin{document}

\section{INTRODUCTION}
The solution of high-dimensional, multiscale system is challenging as the required computational resources usually grow exponentially with the dimension of the state-space as well as with the smallest time-scale that needs to be resolved. As such systems are ubiqitious in applied physics and engineering, reduced/coarse-grained descriptions and models are necessary that are predictive of various observables or the high-dimensional system, but whose discretization time-scales can be much larger than the inherent ones \cite{givon_2004}.\\
\indent
We adopt a data-based perspective \cite{gharamani_2015,lecun_2015} that relies on data generated by simulations of a fine-grained (FG) system in order to learn a coarse-grained (CG) model.
We nevertheless note that such coarse-graining tasks exhibit fundamental differences from large-scale machine learning tasks \cite{koutsourelakis_2016,alber} as the data involved is usually small due to the expensive data acquisition and as information about the underlying physical structure of the problem is available. When this domain knowledge is 
 incorporated into the CG model
  it can improve its predictive ability 
  \cite{kaltenbach_2020, stinis_2019}.\\
\indent
In contrast to other  frameworks for reduced-order modeling (e.g.  SINDy \cite{brunton2016}) where the dynamics of the CG model is learned based on a large vocabulary of feature functions, we employ a deep neural network for the CG dynamics in order to gain great flexibility and be able to not restrict ourselves to an a priori chosen set of feature functions. This approach is similar to the ideas of Neural ODEs \cite{chen} and Neural SDEs \cite{sode_2020} which also use neural networks to represent the dynamics. Another possibility would be the use of Gaussian Processes \cite{raissi} which would allow non-parametric, probabilistic modeling.\\
\indent
In this paper, we combine a generative, probabilistic machine learning framework \cite{koutsourelakis_2011} with virtual observables \cite{kaltenbach_2020} and deep neural networks for the CG dynamics as well as the mapping from the CG states to the FG states. In doing so, we propose a framework that can make use of the flexibility of neural nets, while still obeying physical laws. We carry out the tasks of model estimation and dimensionality reduction simultaneously and identify the CG states variables, their dynamics as well as a  probabilistic coarse-to-fine map based only on small amounts of FG simulation data. 

\section{METHODOLOGY}
In general, the subscript $f$ or lower-case letters are used to denote variables associated with the (high-dimensional) fine-grained(FG) model and the subscript $c$ or upper-case letters are used for quantities of the (lower-dimensional) coarse-grained(CG) description. We also use a circumflex \textbf{$~\hat{}~$} to denote observed/known variables.

\subsection{The FG and CG model}
The fine-grained system considered is a high-dimensional system with state variables $\bx$ ($\bx \in \mathcal{X}_f \subset  \mathbf{R}^{d_f}$), whose dimension $d_f$ is very large ($d_f>>1$). We describe the dynamics of such a FG system by a system of deterministic or stochastic ODEs i.e., 
\vskip-0.6cm
\begin{eqnarray}
\dot{\bx_t} = \bs{f}(\bx_t,t), \quad t>0    
\label{eq:fg}
\end{eqnarray}
The FG system is moreover considered to have the initial condition  $\bx_0$ that might be deterministic or drawn from a specified distribution. In this work, we want to coarse-grain such a system only based on simulated data, i.e. time sequences simulated from \refeq{eq:fg} with a time-step $\delta t$.\\
\indent
Our goal is to simultaneously identify (unknown) CG state variables $\bxx$  with   $\bxx \in \mathcal{X}_c \subset  \mathbf{R}^{d_c} $ as well as the dynamics of those CG variables. The dimension $d_c$ of these CG state variables is intended to be much smaller than $d_f$. For the CG dynamics a Markovian dynamic is assumed in the form:
\vskip-0.6cm
\begin{eqnarray}
\dot{\bxx_t} = \bs{F}(\bxx_t,t), \quad t>0  
\label{eq:cg}
\end{eqnarray}
\subsection{Emission law}
\label{sec:emission}
In contrast to approaches based on the Mori-Zwanzig formalism \cite{mori, zwanzig}, which include a mapping from the FG system to the quantities of interest, we employ a probabilistic, generative coarse-to-fine map \cite{schoberl2017} from the CG state-variables to the FG description. We indicate the associated (conditional) density by:
\vskip-0.6cm
\begin{eqnarray}
p_{cf}(\bx_t |~\bxx_t; ~\bt_{cf})
\label{eq:pcf}
\end{eqnarray}
where $\bt_{cf}$ denote the (unknown) parameters that we will try to learn from the data.
This conditional density $p_{cf}$ can be endowed a priori with domain knowledge by adapting its form to the particulars of the problem or it can parametrized by deep neural networks to allow for maximum flexibility.\\
\indent
Employing a probabilistic coarse-to-fine map instead of a deterministic, restriction operator has many advantages as  e.g. the full FG system's reconstruction and probabilistic predictive estimates. 

\subsection{Transition law}
\label{sec:transition}
In the following, we consider discretized time with a fixed time-step $\Delta t$ and time-related subscripts refer to  the number of time-steps.\\
\indent
We model the CG dynamics with the help of a deep neural network in order to gain a great flexibility and be able to express nonlinear functions. Therefore, we assume an explicit discretization of \refeq{eq:cg} and model the right-hand-side by the deep neural network $NN(.)$ parametrized by  $\bt_{NN}$:
\vskip-0.6cm
\begin{eqnarray}
\bxx_{t+1} = \bxx_{t}+NN(\bxx_{t},\bt_{NN}) + \sigma_{r} \boldsymbol{\epsilon}, \qquad \boldsymbol{\epsilon} \sim \mathcal{N}(\bs{0}, \bs{I})
\label{eq:cg_dis}
\end{eqnarray}
where the parameter $\sigma_r \geq 0$ is responsible for the stochastic part of the CG dynamics. This leads to the following conditional density:
\vskip-0.6cm
\begin{eqnarray}
 p(\bxx_{t+1}~ | \bxx_{t}, \bt_{NN}, \sigma_{r}) = \mathcal{N}(\bxx_{t+1} ~| ~\bxx_{t}+NN(\bxx_{t},\bt_{NN}), \sigma^2_{r} \bs{I})
 \label{eq:vtransition}
 \end{eqnarray}
  which effectively   represents a discretized version of the neural stochastic ODEs of \cite{sode_2020} and is more flexible as compared to  approaches in which the  right-hand side consists of a restricted amount of first- and second-order interactions of $\bxx_t$ \cite{kaltenbach_2020}.

\subsection{Virtual observables}
As the CG state-variables $\bxx$ employed in multiscale modeling are usually given physical meaning, we employ the concept of {\em virtual observables} \cite{kaltenbach_2020} in order to incorporate general physical principles such as conservation of mass, momentum or energy. Let these be expressed   as equalities of the form at each time-step $l$:
\vskip-0.6cm
\begin{eqnarray}
 \bs{c}_l(\bxx_{l }) = \bs{0}, \quad l=0,1,\ldots
 \label{eq:constraintl}
 \end{eqnarray}
where  $\bs{c}_l: \mathcal{X}_c \subset \RR^{d_c} \to \RR^{M_c}$. 
The only requirement we will impose is that of differentiability of $\bs{c}_l$ \cite{kaltenbach_2020}.
We define a new variable $\hat{\bs{c}}_l$ which relates to  $\bs{c}_l$ as follows:
\vskip-0.6cm
\begin{eqnarray}
 \hat{\bs{c}}_l = \bs{c}_l (\bxx_l) +\sigma_c \beps_c, \qquad \beps_c \sim \mathcal{N}(\bs{0}, \bs{I})
 \label{eq:vocon}
 \end{eqnarray}
 Now, it is assumed that the $\hat{\bs{c}}_l$ have been {\em virtually} observed and this set of virtual observations $\hat{\bs{c}}_l=0$ leads to to an augmented version of the FG data and therefore virtual likelihoods of the type:
\vskip-0.6cm
\begin{eqnarray}
 p(\hat{\bs{c}}_l=\bs{0}~ | ~\bxx_l, \sigma_R) = \mathcal{N}(\bs{0} ~| ~\bs{c}_l (\bxx_l), \sigma^2_c \bs{I})
 \label{eq:vlikeres}
 \end{eqnarray}
 The ``noise" parameter $\sigma_c$ can be used to account for the intensity of the enforcement of the virtual observations and represents the tolerance parameter with which the constraints would be enforced in a deterministic setting.\\
 \indent
 We note that the concept of virtual observables is not restricted to physical constraints but could also be applied to residuals of temporal discretization schemes \cite{kaltenbach_2020} or of PDEs \cite{rixner_2020}. In both of this cases, it is shown that the incorporation of virtual observables can reduce the amount of training data required and  enable training in the Small Data regime.
 
\subsection{Inference and learning}
\label{sec:inference}
Due to the introduction of virtual observables, we can adopt an enlarged definition of data which we cumulatively denote by $\mathcal{D}=\left \{ \hat{\bx}_{ 0: T }^{(1:n)},\hat{\bs{c}}_{0:T}^{(1:n)}\right \}$ and which encompasses:
\bi
\item FG simulation data consisting of $n$ sequences of the FG state-variables. These are denoted by $\hat{\bx}_{ 0: T }^{(1:n)}$ as the likelihood model implied by the $p_{cf}$ in \refeq{eq:pcf} involves only the observables at each coarse time-step.

\item Virtual observables $\hat{\bs{c}}_l^{(1:n)}$ relating to the CG states $\bxx_l$  at each time-step $l$ and which relate to the physical constraints 
as in \refeq{eq:vocon}. In the example they pertain to all time-steps from $0$ to $T$ and are denoted by $ \hat{\bs{c}}_{0:T}^{(1:n)}$.
\ei
\indent
We represent the latent (unobserved) variables of the model by the CG state-variables $\bxx_{ 0: T }^{(1:n)} $  and relate them to the FG data through the $p_{cf}$ (in \refeq{eq:pcf}) and to the virtual observables through \refeq{eq:vlikeres}. 
The parameters of the model are denoted cumulatively by $\bt$ and consist of\footnote{If any of these parameters are prescribed, then they are omitted from $\bt$.}:
\bi
\item $\bt_{NN}$ which parametrize the neural network for the right-hand-side of the CG evolution law (see section \ref{sec:transition}),
\item  $\bt_{cf}$ which parametrize the probabilistic coarse-to-fine map (\refeq{eq:pcf}),
\item $\sigma_r$ involved in the stochasticity of the transition law \refeq{eq:cg_dis} and
\item $\sigma_c$ involved in the enforcement of virtual observables in \refeq{eq:vocon}
\ei

We follow a fully-Bayesian formulation and express the posterior of the unknowns (i.e. latent variables and parameters) as follows:
\vskip-0.6cm
\begin{eqnarray}
p( \bxx_{ 0: T }^{(1:n)}, ~\bt~ |~ \mathcal{D}) = \cfrac{ p(\mathcal{D}~ | ~\bxx_{ 0: T }^{(1:n)}, \bt) ~p(\bxx_{ 0: T }^{(1:n)}, \bt) }{p(\mathcal{D}) }
\label{eq:genbayes}
\end{eqnarray}
where $p(\bxx_{ 0: T }^{(1:n)}, \bt)$ denotes the prior on the latent variables and parameters. The likelihood term $p(\mathcal{D} | \bxx_{ 0: T }^{(1:n)}, \bt)$ involved can be decomposed into the product of two (conditionally) independent terms, one for the FG data and one for the virtual observables, i.e.:
\vskip-0.6cm
\begin{eqnarray}
p(\mathcal{D} ~|~ \bxx_{ 0: T }^{(1:n)}, \bt) = p(\hat{\bx}_{ 0: T }^{(1:n)} ~|~  \bxx_{ 0: T }^{(1:n)}, \bt) ~ p(\hat{\bs{c}}_{0:T}^{(1:n)} ~|~  \bxx_{ 0: T }^{(1:n)}, \bt)
\label{eq:likedecomp}
\end{eqnarray}
We further note that (from \refeq{eq:pcf}):
\vskip-0.6cm
\begin{eqnarray}
p(\hat{\bx}_{ 0: T }^{(1:n)}~ | ~ \bxx_{ 0: T }^{(1:n)}, \bt)  = \prod_{i=1}^n \prod_{t=0}^T p_{cf}( \bx^{(i)}_{t~} ~|~  \bxx^{(i)}_{t~}, \bt_{cf})
\label{eq:fgdatalikelihood}
\end{eqnarray}
and (from \refeq{eq:vlikeres}):
\vskip-0.6cm
\begin{eqnarray}
\begin{array}{ll}
p( \hat{\bs{c}}_{0:T}^{(1:n)}  |  \bxx_{ 0: T }^{(1:n)}, \bt)  &  = \prod_{i=1}^n \prod_{l=0}^T \mathcal{N}(\bs{0} | \bs{c}_l( \bxx^{(i)}_{l}), \sigma_c^2 \bs{I}) \\
& \propto \prod_{i=1}^n \prod_{l=0}^T  \frac{1}{\sigma_c^{dim(\bs{c})} } \exp \left\{ -\frac{1}{2 \sigma_c^2} \left| \bs{c}_l( \bxx^{(i)}_{l~})  \right|^2 \right\}
\end{array}
\end{eqnarray}
The prior $p(\bxx_{ 0: T }^{(1:n)}, \bt)$ can be decomposed into the transition density of \refeq{eq:vtransition} and a prior for $\bxx_0$ as well as the parameters $\bt$:
\vskip-0.6cm
\begin{eqnarray}
p(\bxx_{ 0: T }^{(1:n)}, \bt)    = \prod_{i=1}^n \; p(\bxx_0^{(i)}) \prod_{t=0}^{T-1} p(\bxx_{t+1}^{(i)} ~| ~\bxx_{t}^{(i)},\bt_{NN}, \sigma_{r}) \;  \; p(\bt) 
\end{eqnarray}
We advocate the use of Stochastic Variational Inference \cite{svi} for computing an approximate posterior. We select a  parameterized family of densities, $q_{\bs{\phi}}( \bxx_{ 0: T }^{(1:n)}, ~\bt)$ and  attempt to find the one that best approximates the posterior by minimizing their Kullback-Leibler divergence. It can be shown \cite{bishop}, that this optimal $q_{\bs{\phi}}$  maximizes the Evidence Lower Bound (ELBO) $\mathcal{F}(q_{\bp}( \bxx_{ 0: T }^{(1:n)}, ~\bt))$:
\vskip-0.6cm
\begin{eqnarray}
\begin{array}{ll}
 \log p(\mathcal{D}) & =\log  \int p( \mathcal{D}, ~\bxx_{ 0: T }^{(1:n)}, ~\bt ) ~d\bxx_{ 0: T }^{(1:n)} ~d\bt \\
 & = \log  \int \cfrac{ p( \mathcal{D} | ~\bxx_{ 0: T }^{(1:n)}, ~\bt ) p( \bxx_{ 0: T }^{(1:n)}, ~\bt )}{ q_{\bp}( \bxx_{ 0: T }^{(1:n)}, ~\bt)} q_{\bp}( \bxx_{ 0: T }^{(1:n)}, ~\bt) ~d\bxx_{ 0: T }^{(1:n)} ~d\bt \\
 & \ge \int \log \cfrac{p( \mathcal{D} | ~\bxx_{ 0: T }^{(1:n)}, ~\bt ) p( \bxx_{ 0: T}^{(1:n)}, ~\bt )}{ q_{\bp}( \bxx_{ 0: T }^{(1:n)}, ~\bt)} q_{\bp}( \bxx_{ 0: T }^{(1:n)}, ~\bt) ~d\bxx_{ 0: T }^{(1:n)} ~d\bt \\
 & = \mathcal{F}(q_{\bp}( \bxx_{ 0: T }^{(1:n)}, ~\bt))
\end{array}
\label{eq:elboqphi}
\end{eqnarray}
\noindent
In the following illustrations, we postulate a {\em mean-field}  decomposition:
\vskip-0.6cm
\begin{eqnarray}
q_{\bp}( \bxx_{ 0: T }^{(1:n)}, ~\bt)  = q_{\bp}( \bxx_{ 0: T}^{(1:n)}) ~ p_{\bp}(\bt)   = \left[ \prod_{i=1}^{n} q_{\bp}( \bxx_{ 0: T }^{(i)}) \right]~~ \delta_{\bp}(\bt)
\label{eq:qfactor}
\end{eqnarray}
where we make  use of 
the (conditional) independence of the time sequences in the likelihood. We further note that we employed Dirac $\delta_{\bp}$ functions for the $q_{\bp}(\bt)$ and therefore obtain MAP estimates $\bt_{MAP}$ (i.e. $\bp$ includes $\bt_{MAP}$) for the unknown parameters.\\
\indent
Gradients of the ELBO with respect to the parameters $\bp$  
involve expectations with respect to $q_{\bp}$. These were approximated with Monte Carlo estimates which employ the reparametrization trick \cite{kingma_2014} and stochastic optimization was carried out with the ADAM algorithm \cite{adam}.

\subsection{Predictions}
The proposed framework can produce probabilistic predictive estimates for a sequence which was  observed up to time-step $T$ i.e. $\hat{\bx}^{(i)}_{0:T}$. This predictive uncertainty reflects not only the information-loss due to the coarse-graining process but also the epistemic uncertainty arising from  finite (and small) datasets.\\
\indent
In particular, if $q_{\bp}(\bxx_{T}^{(i)})$ is the (marginal) posterior of the last,  hidden CG state and $\bt_{MAP}$ the MAP estimate of the model parameters, then we follow the steps described in Algorithm \ref{alg:preda}. This procedure generates samples of the full FG state evolution but does not necessarily guarantees the enforcement of the constraints for the CG states.\\
\indent
We note that if we would also like to enforce the constraints $\bs{c}_l$ for future predictions, then these would need to be included in  the posterior density defined in \refeq{eq:genbayes}. Consequently,  future (FG or CG) states would need to be inferred from this augmented posterior and an enlarged inference process is required for predictions.
\begin{algorithm}[!ht]
\SetAlgoLined
\KwResult{Sample of $\bx_{ (T+P)}^{(i)}$}
\KwData{$q_{\bp}(\bxx_{T}), \bt_{MAP}$}
 Sample from $q_{\bp}(\bxx_{T}^{(i)})$\;
 \While{Time-step $(T+P)$  not reached}{
  Sample from the CG evolution law  in \refeq{eq:cg_dis}\;
 }
 Sample from $p_{cf}(\bx_{(T+P)}~|~ \bxx_{(T+P)}, \bt_{MAP})$
 \caption{ Prediction - Algorithm }
 \label{alg:preda}
\end{algorithm}

\newpage
\section{NUMERICAL ILLUSTRATIONS}
We demonstrate the capabilities of the proposed framework by applying it to a high-dimensional system of stochastically moving particles.
\subsection{FG model}
For the simulations presented in this section, we used  $d_f=250 \times 10^3$ particles, which, at each microscopic time step $\delta t=2.5 \times 10^{-3}$  performed random, non-interacting,  jumps of size $\delta s=\frac{1}{640}$, either  to the left with probability $p_{left}=0.1875$ or to the right with probability $p_{right}=0.2125$. The positions were restricted to a domain of $[-1,1]$ with periodic boundary conditions.  It is well-known \cite{cottet} that in the limit (i.e.  $d_f \to \infty$) 
 the particle density $\rho(s,t)$ can be described with an advection-diffusion PDE with diffusion constant $D=(p_{left}+p_{right})\frac{\delta s^2} {2\delta t}$ and velocity $v=(p_{right}-p_{left})\frac{\delta s}{\delta t}$:
  \vskip-0.6cm
\begin{eqnarray}
  \cfrac{\pa \rho }{\pa t} +v \cfrac{\pa \rho}{\pa s}=D \frac{\pa^2 \rho}{\pa s^2}, \qquad s \in (-1,1)..
  \label{eq:addensity}
  \end{eqnarray}

\subsection{CG model specifications}
The CG model relates to a discretization of the particle density into $d_c=25$ equally-sized bins at each coarse time step .
The nature of the CG variables $\bxx_t$ gives rise to  a multinomial for the coarse-to-fine density $p_{cf}$ (section \ref{sec:emission}) i.e.:\\
\begin{equation}
p_{cf}(\bs{x}_t | \bs{X}_t)=  \frac{ d_f !}{m_1(\bx_t)!~m_2(\bx_t)! \ldots m_{d_c}(\bx_t)!} \prod_{j=1}^{d_c}X_{t,j}^{m_j(\bx_t)} 
 , \quad \textrm{} 
 \label{eq:Coarse_to_fine_particle}
\end{equation}
where $m_j(\bx_t)$ is the number of particles in bin $j$. We assume that, given the CG state $\bxx_t$, the coordinates of the particles $\bx_t$ are  conditionally independent. This does not imply that they move independently nor that they cannot exhibit coherent behavior \cite{felsberger_2019}.
The consequence of \refeq{eq:Coarse_to_fine_particle} is that for this example no parameters need to be learned for $p_{cf}$.\\
\indent
For the transition law (section \ref{sec:transition}), we assume a coarse time step of $\Delta t=4$ and employed a two-layered fully connected neural network $NN(.)$ with ReLU activation functions. Each layers consisted of 25 neurons.
We enforce  conservation of mass, using the following constraint at each time step $l$:
\vskip-0.6cm
\begin{eqnarray}
 c_l(\bxx_{l }) = 1- \sum_{j=1}^{d_c} X_{l,j} =0, \quad l=0,1,\ldots
 \label{eq:constraint_ex}
 \end{eqnarray}
These are complemented by the virtual observables presented earlier and with  $\sigma_c^2=10^{-9}$ (\refeq{eq:vocon}).\\
\indent
For the family of variational distributions $q_{\bp}(\bxx^{(i)}_{(0:T)})$ and since $X_{t,j}^{(i)}>0, \forall j,t$, we employed  multivariate lognormals  with a diagonal covariance matrices i.e. we assume $X_{t,j}^{(i)}$ are a posteriori independent. The mean and covariance matrix of the underlying Gaussians for each sequence $i$ become part of the parameters $\bp$ with respect to which the ELBO is maximized (see Section \ref{sec:inference}). 
We note that it would also be possible to use an amortized formulation and explicitly account for the dependence on the data values by employing a neural network for both mean and covariance with the time sequence as an input.

\subsection{Results}
\begin{figure}[!ht]
 \centering
\includegraphics[width=0.7\textwidth,trim=0 20 0 60,clip]{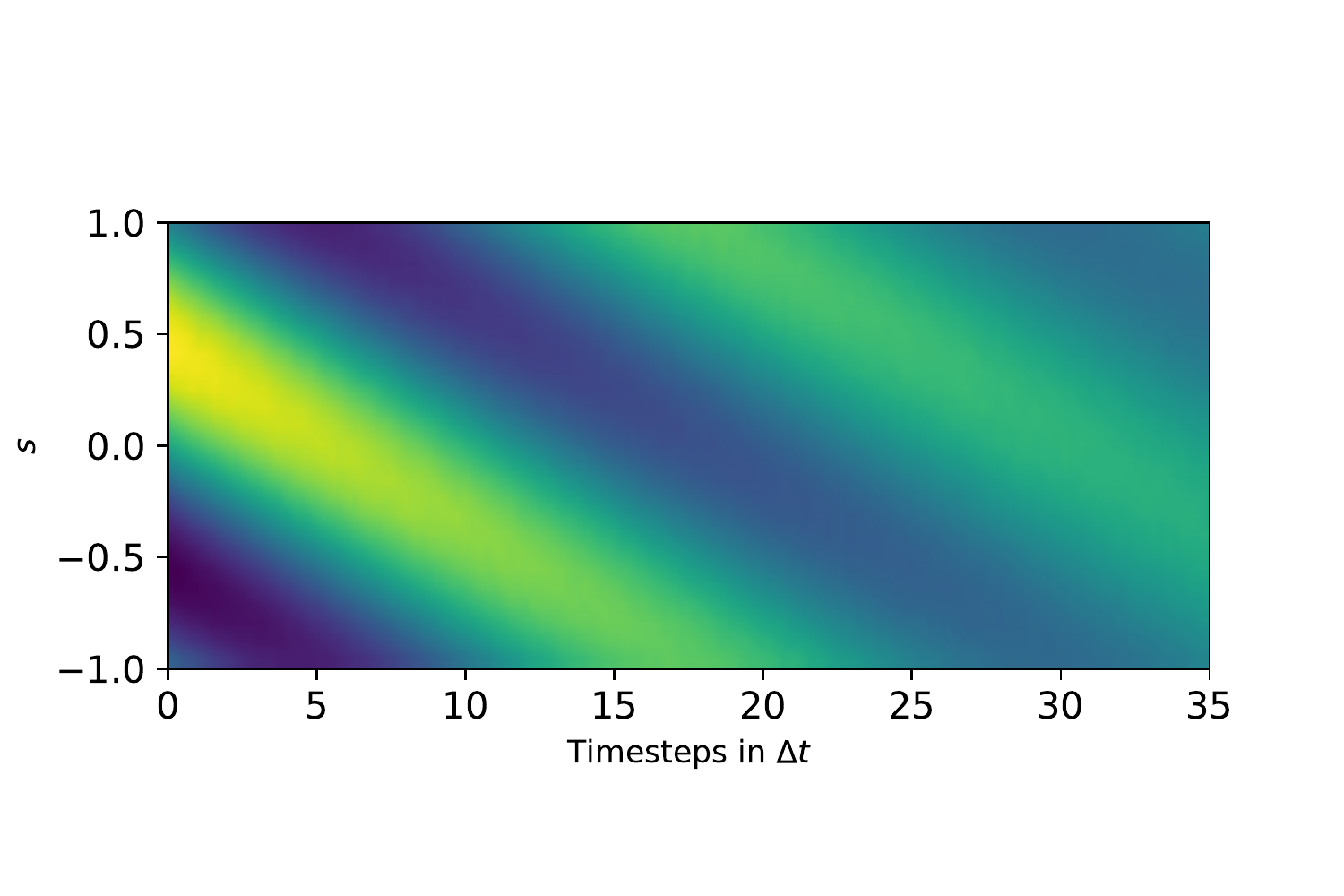}\\
\includegraphics[width=0.7\textwidth,trim=0 20 0 60,clip]{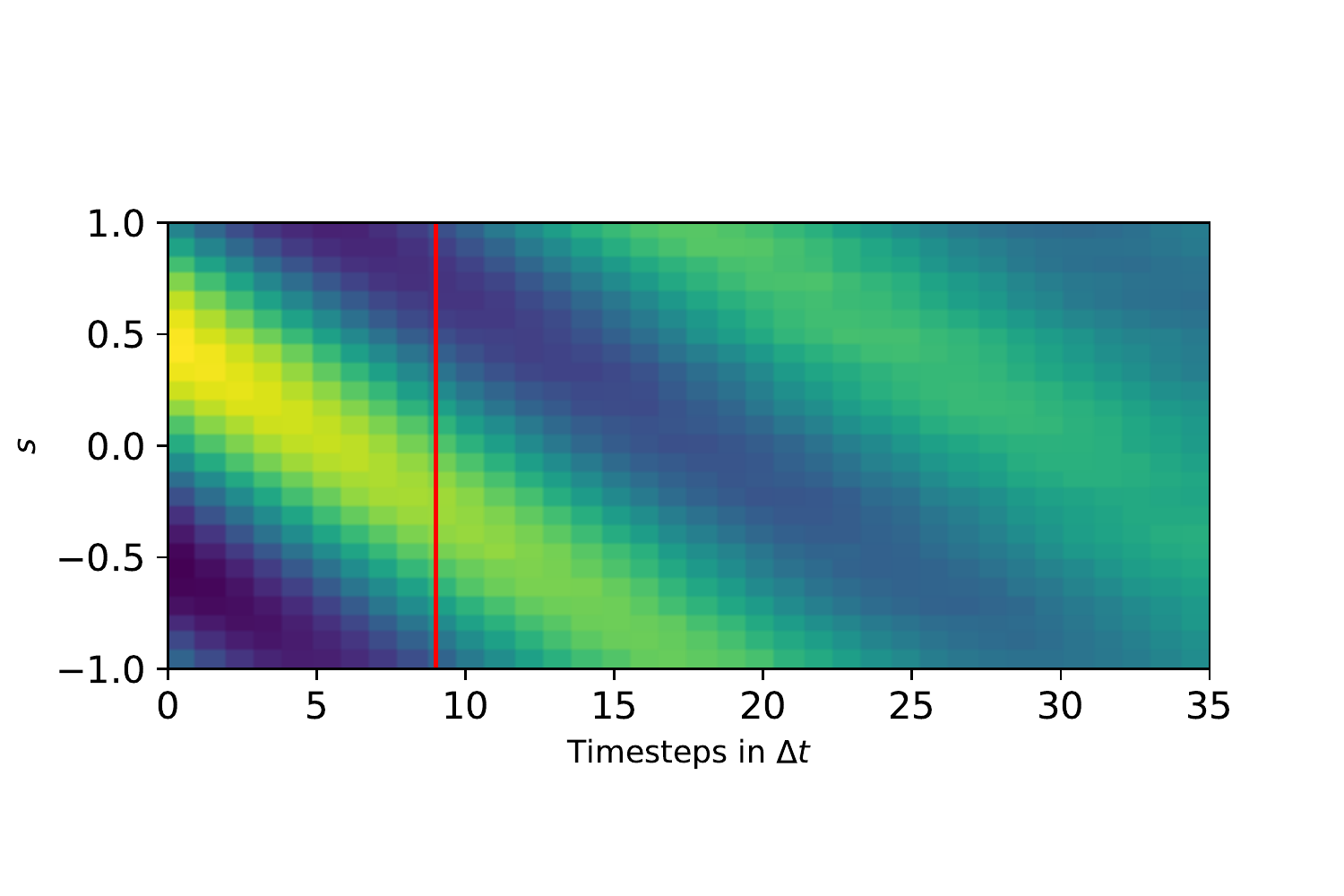}
\caption{Particle density: Inferred and predicted posterior mean (bottom) in comparison with the ground truth (top). The red line divides inferred quantities from predicted ones.}
\label{fig:ad_pred_large}
\end{figure}

We employed $n=64$ time sequences with $T=9$ 
 for training and  applied our framework in order to infer the unobserved CG states but more importantly the model parameters in right-hand side of the CG dynamics.\\
\indent
In Figure 1 we compare the true particle density with the one predicted by the trained CG model for one illustrative time sequence. We note that the latter is computed by reconstructing the $x_t$ futures. The trained model is able to accurately track first-order statistics well into the future  for many more time steps than those contained in the training data.\\
\indent
A more detailed view of the predictive estimates with snapshots of the particle density at selected time instances is presented in Figure 2 and 3 where the predictive posterior mean but also the associated uncertainty is displayed. Inferred as well as predicted particle densities match accurately the ground-truth and reasonable uncertainty bounds are computed. 

\begin{figure}[h]
\centering
\includegraphics[width=0.32\textwidth]{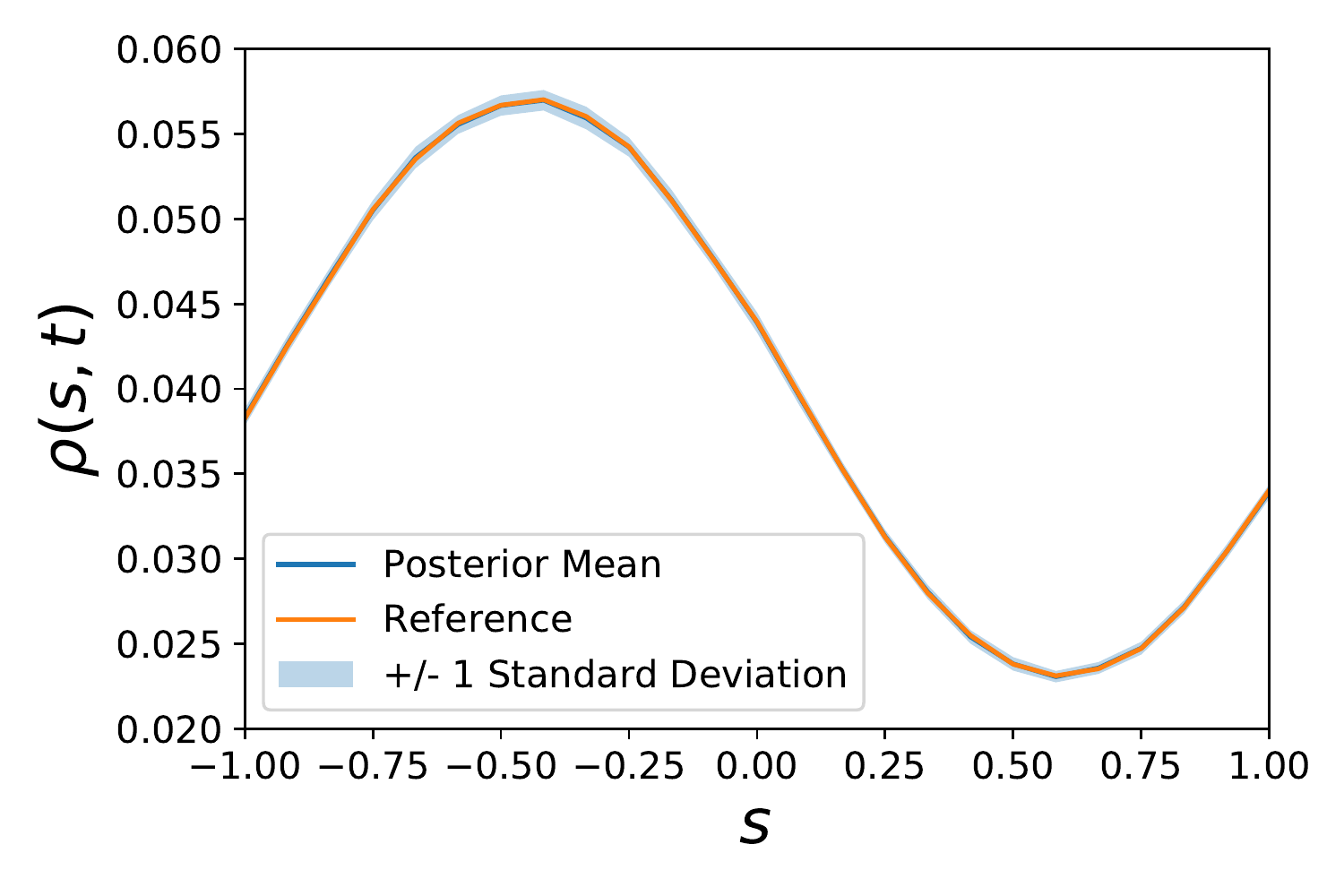}
\includegraphics[width=0.32\textwidth]{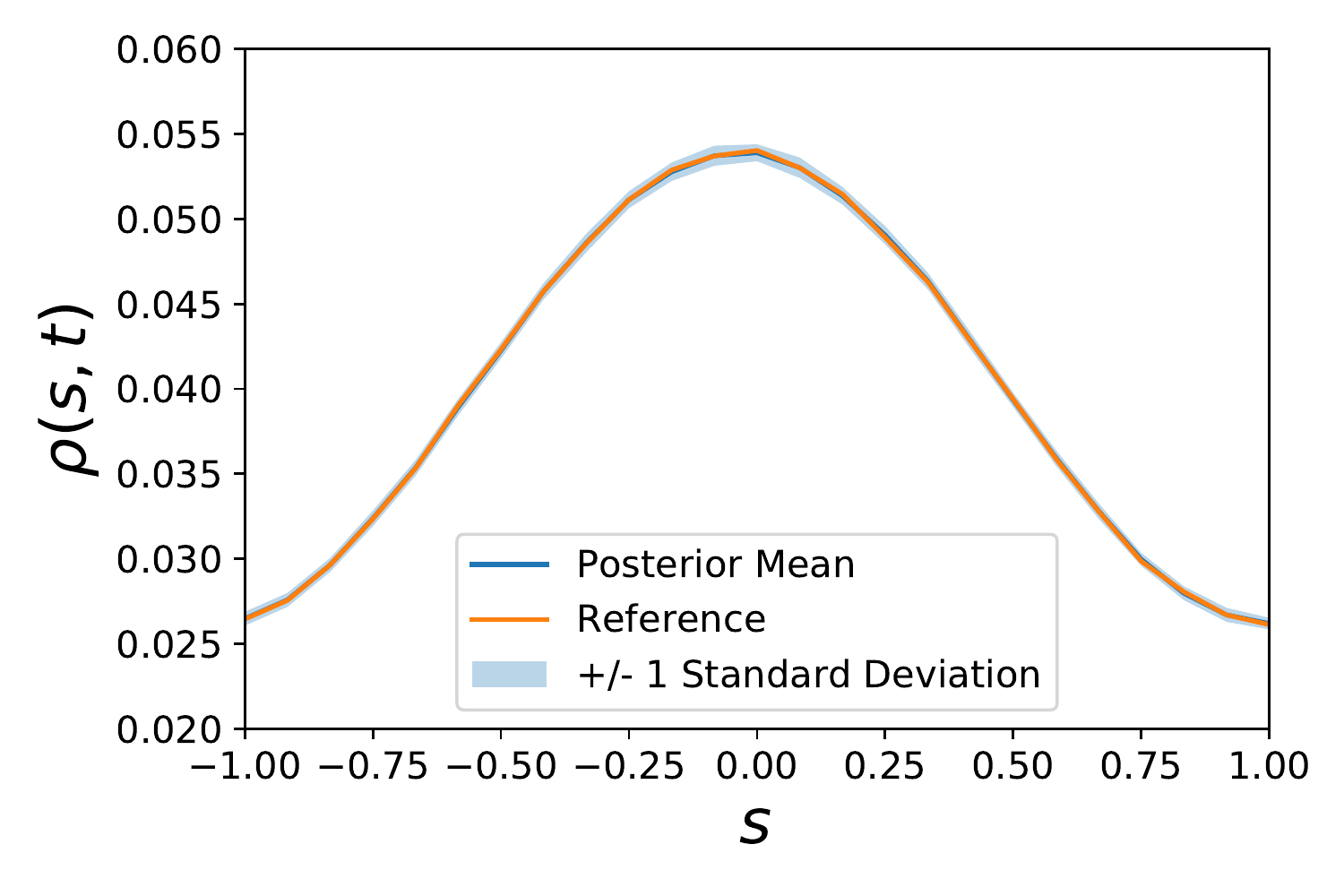}
\includegraphics[width=0.32\textwidth]{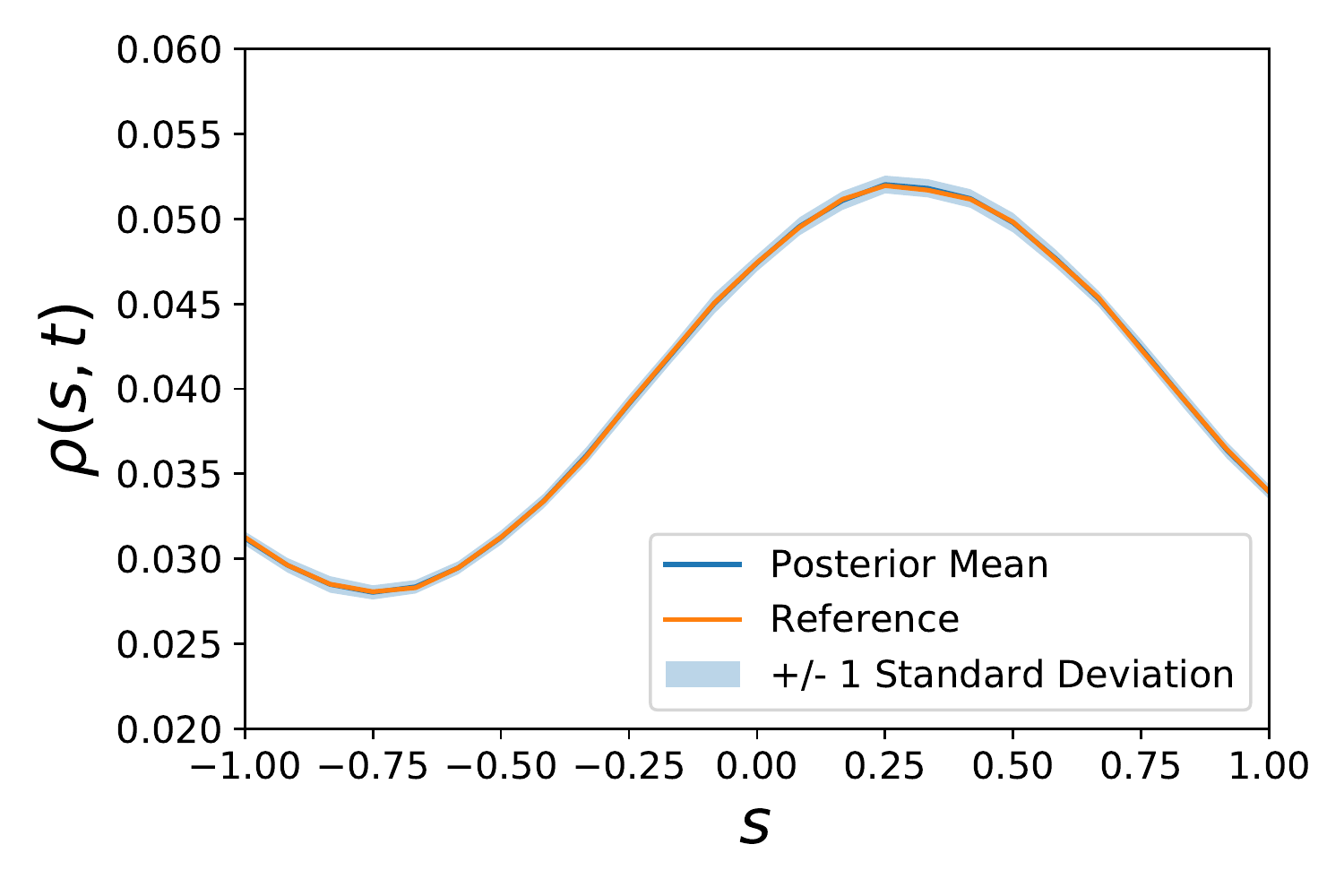}
\caption{Inferred particle density profiles  at $t=0,5 \Delta t,9 \Delta t$ (from left to right).}
\label{fig:ad_inf}
\end{figure}

\begin{figure}[h]
\centering
\includegraphics[width=0.4\textwidth]{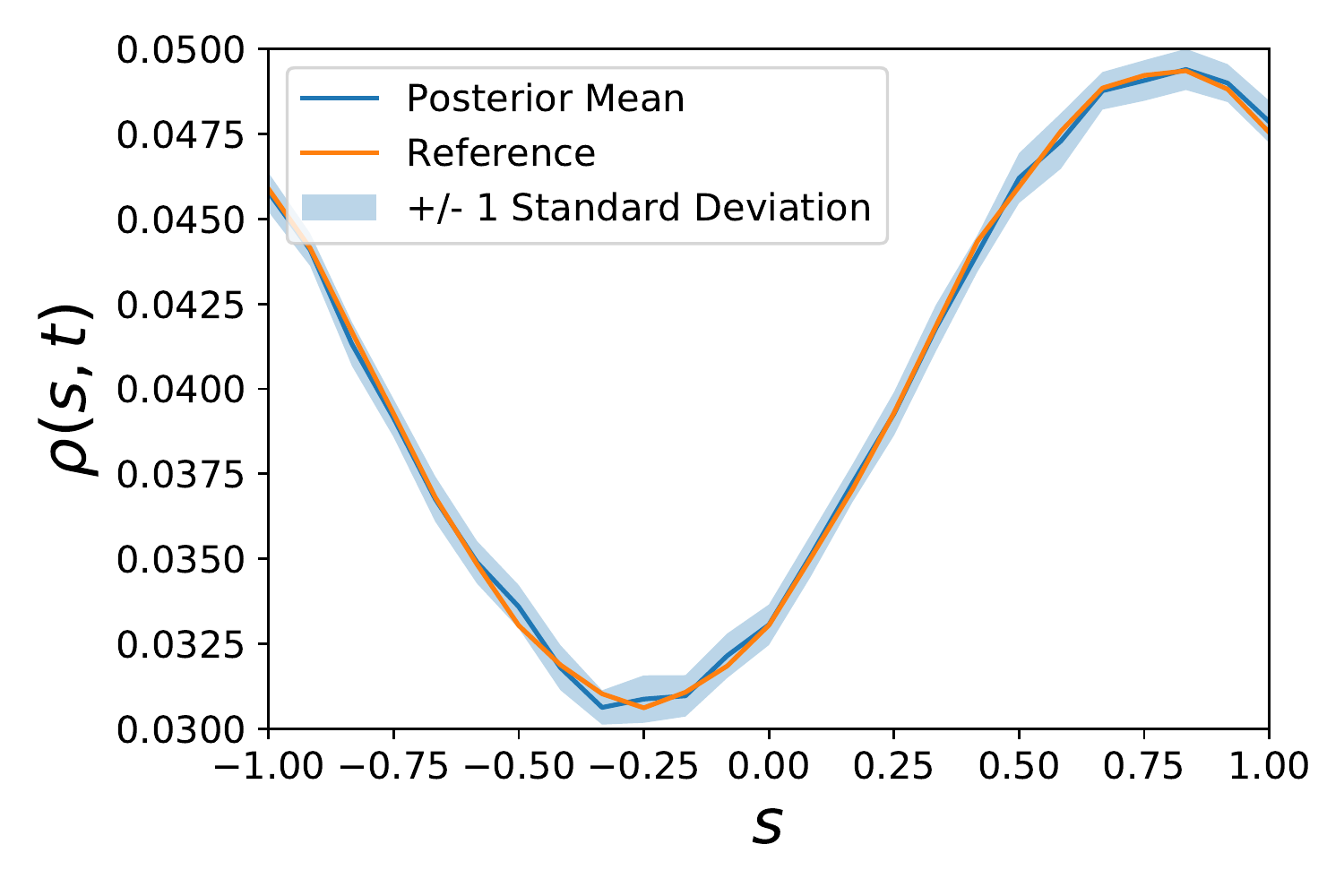}
\includegraphics[width=0.4\textwidth]{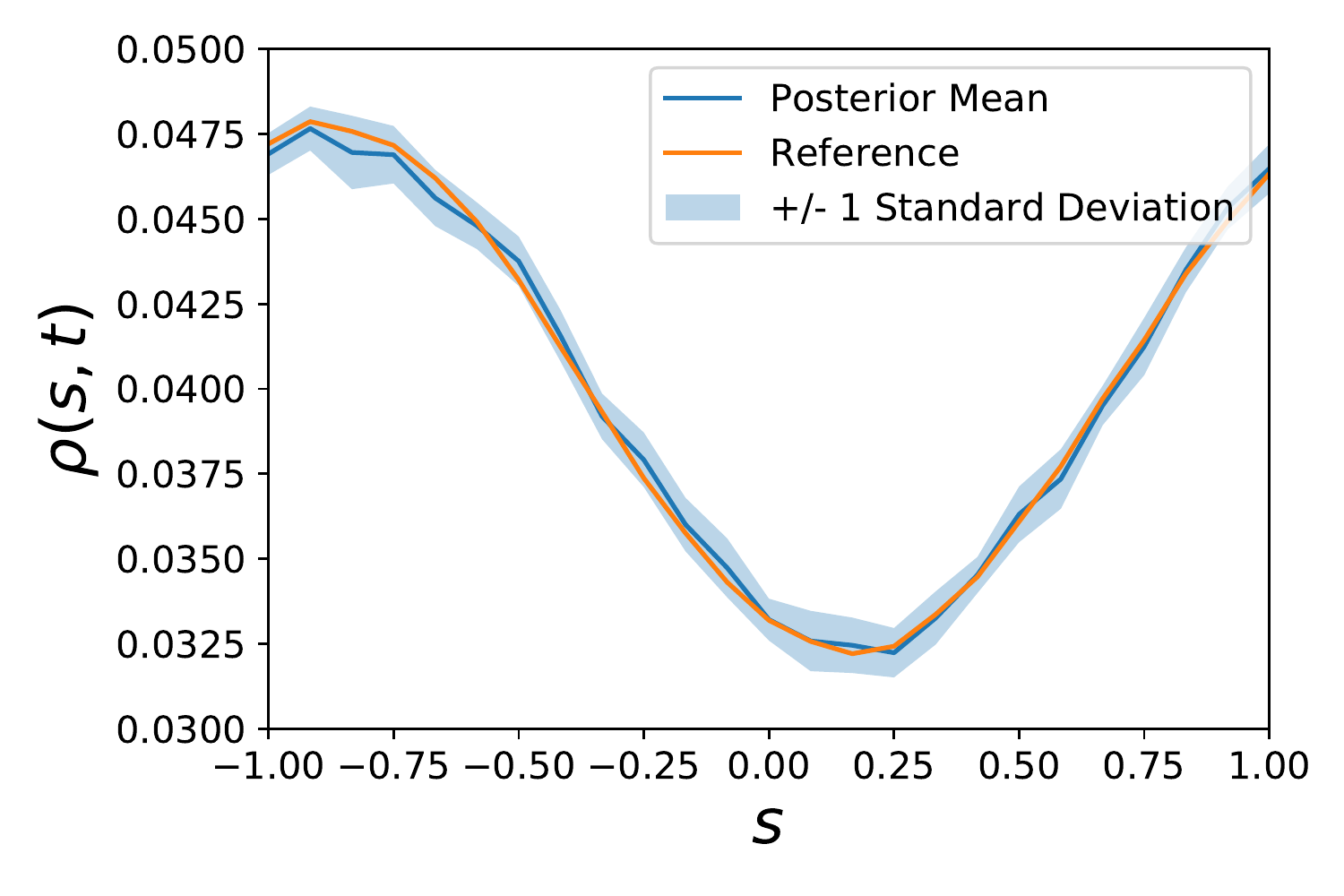}\\
\includegraphics[width=0.4\textwidth]{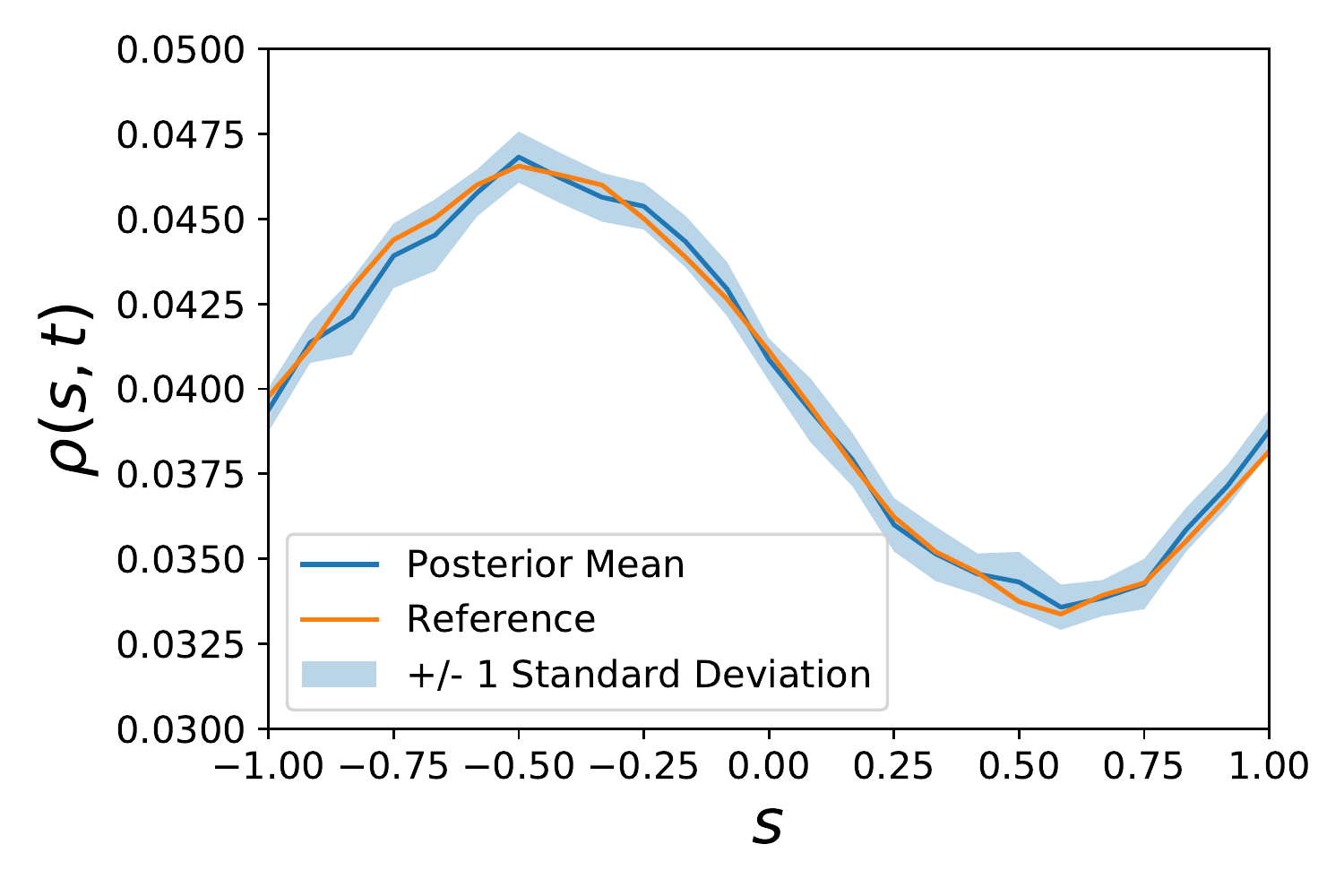}
\includegraphics[width=0.4\textwidth]{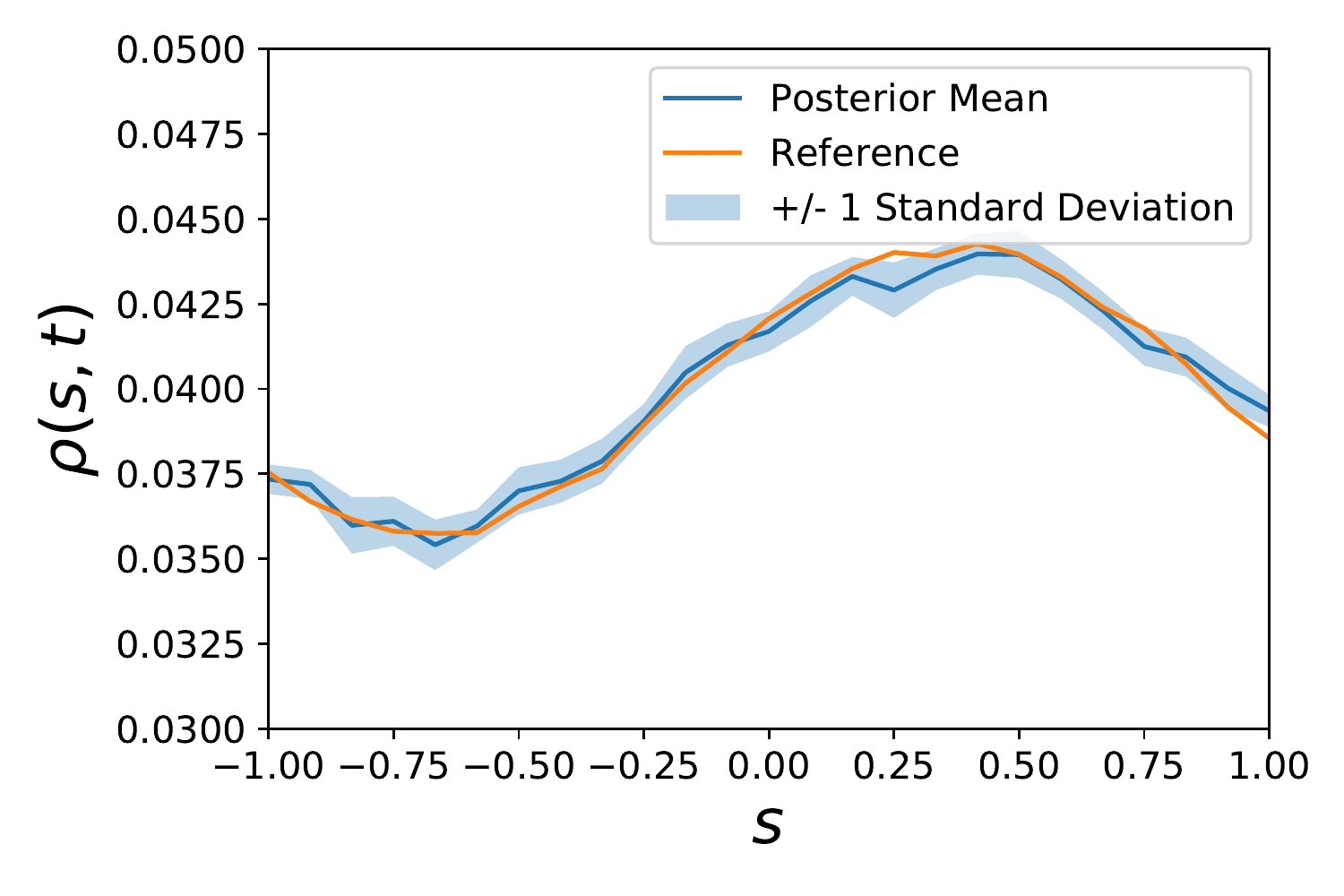}
\caption{Predicted particle density profiles  at $t=15 \Delta t,20 \Delta t,25 \Delta t, 35 \Delta t$ (from left to right and top to bottom).}
\label{fig:ad_pred}
\end{figure}

Finally, in Figure 4,  the mass constraint is depicted for inferred as well as predicted particle densities and good agreement with the target value ($=1$) is observed. This result is particularly important as it demonstrates that the virtual observables were able to find $CG$ state variables that agree with an a priori given physical constraint and additionally a transition law has been learned that is able to automatically  satisfy the constraint in the future.

\begin{figure}[!ht]
\centering
\includegraphics[width=0.5\textwidth]{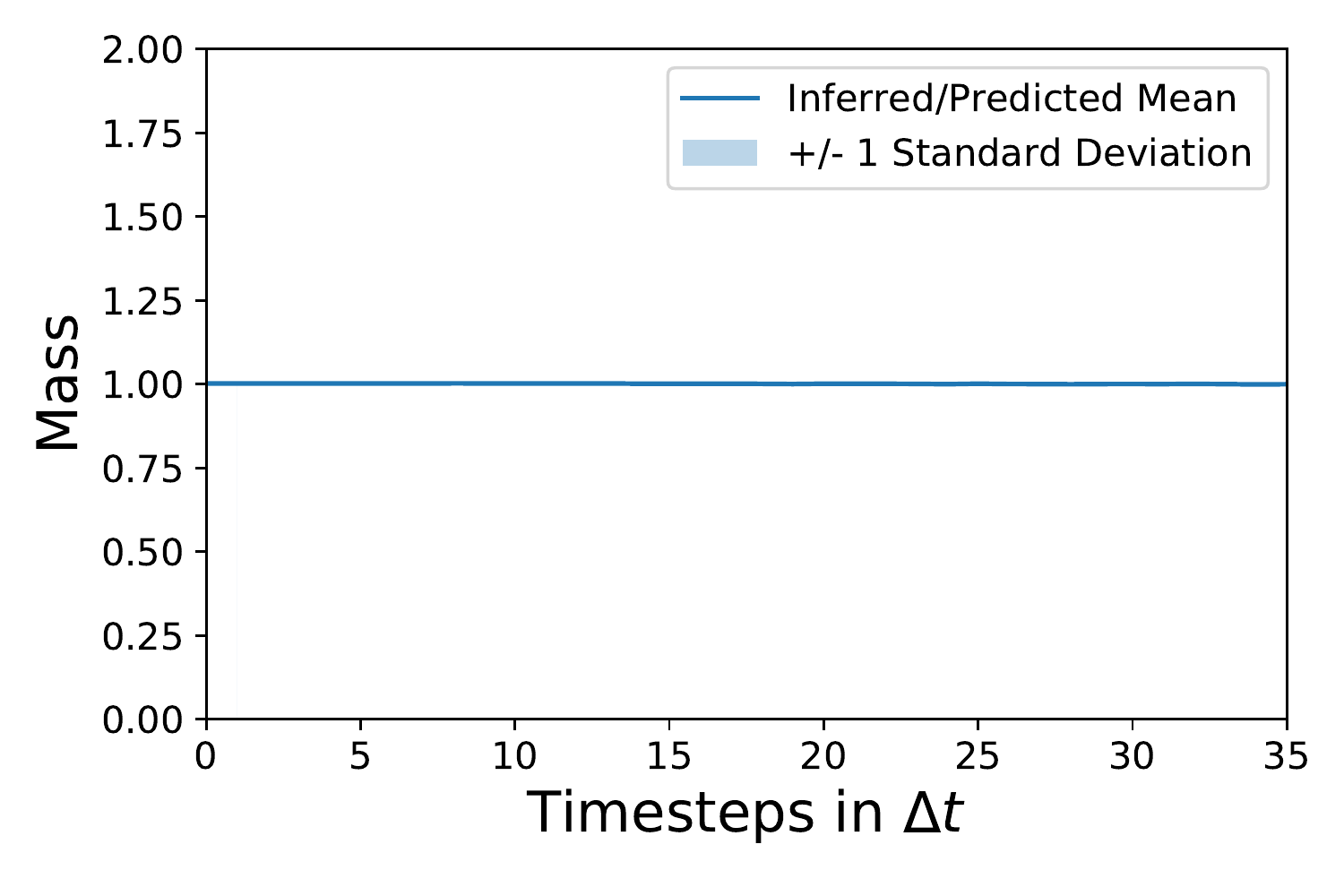}
\caption{Mass based on inferred and predicted particle densities.}
\label{fig:ad_mass}
\end{figure}

\section{CONCLUSIONS}
We combined a probabilistic generative model with physical constraints  and deep neural networks in order to obtain a framework for the automated discovery of coarse-grained variables and dynamics based on fine-grained simulation data. The FG simulation data are augmented in a fully Bayesian fashion by virtual observables that enable the incorporation of physical constraints at the CG level. These could be for instance conservation laws that are available when CG variables have physical meaning. Deviations from such conservation laws would  invalidate predictions. 
As a result of augmenting the training data with domain knowledge, the model proposed can  learn from  Small Data (i.e. shorter and fewer FG time-sequences) which is a crucial advantage in multiscale settings where the simulation of the FG dynamics is  computationally very expensive.\\ 
\indent
Our approach learns simultaneously a coarse-to-fine mapping and  a transition law for the coarse-grained dynamics by employing probabilistic inference tools for the latent variables and model parameters. Deep neural networks can be used in   both of  these components in order to endow great expressiveness and flexibility. \\
\indent
The model proposed  was successfully tested on a coarse-graining task which involved stochastic particle dynamics. In the example presented, the method was able to accurately predict particle densities at time steps not contained in the training data. Moreover, as it is able to reconstruct the entire FG state vector at any future time instant, it is capable of producing predictions of any FG observable of interest as well as quantify the associated predictive uncertainty.\\
\indent
A  shortcoming of presented framework is that the CG dynamics are not fully interpretable and long-term stability is not guaranteed. These limitations have been addressed in \cite{kaltenbach_2021}  where an  additional layer of latent variables was employed that ensured the discovery of stable CG dynamics but also promoted the identification of slow-varying processes that are most predictive of the system's long-term evolution. 

\end{document}

%% file: definitions.tex
\usepackage{amssymb}
\usepackage{graphicx}
\usepackage{latexsym}
\usepackage{amsmath}
\usepackage{bbold}      
\usepackage[algoruled, linesnumbered]{algorithm2e}

\usepackage{url}
\usepackage{hyperref}
\usepackage{caption}
\usepackage{subcaption}
\usepackage{lineno}

\usepackage{bm}
\usepackage{xcolor}
\usepackage{algorithm2e}

\usepackage{enumerate}
\usepackage{sidecap}
\usepackage{wrapfig}

\usepackage{comment}    
\usepackage{makecell}   

\usepackage{tikz}
\usetikzlibrary{bayesnet}
\usetikzlibrary{calc,fit,arrows,arrows.meta, backgrounds}
\pgfdeclarelayer{background}
\pgfdeclarelayer{foreground}
\pgfsetlayers{background,foreground}

\definecolor{newcolor}{rgb}{.8,.349,.1}

\newcommand{\bx}{\bs{x}}

\newcommand{\bxx}{\bs{X}}

\newcommand{\bt}{\bs{\theta}}

\newcommand{\bp}{\bs{\phi}}

\newcommand{\beps}{\bs{\epsilon}}

\newcommand{\be}{\begin{equation}}
\newcommand{\ee}{\end{equation}}
\newcommand{\bc}{\begin{center}}
\newcommand{\ec}{\end{center}}
\newcommand{\bd}{\begin{description}}
\newcommand{\ed}{\end{description}}
\newcommand{\bi}{\begin{itemize}}
\newcommand{\ei}{\end{itemize}}
\newcommand{\pa}{\partial}
\newcommand{\bs}{\boldsymbol}

\def\RR{ \mathbb R}

\newcommand{\refeq}[1]{Equation (\ref{#1})}

\newcommand{\bmat}{\begin{pmatrix}}
\newcommand{\emat}{\end{pmatrix}}
\newcommand{\bsmat}{\left(\begin{smallmatrix}}
\newcommand{\esmat}{\end{smallmatrix}\right)}
\newcommand{\bes}{\begin{equation}\begin{split}}
\newcommand{\ees}{\end{split}\end{equation}}

\definecolor{TUMGruen}{RGB}{162,173,0} 
\definecolor{TUMOrange}{RGB}{227,114,34} 